\pdfoutput=1
\documentclass[twocolumn,10pt]{article}

\usepackage[letterpaper,margin=2.25cm]{geometry}
\usepackage[T1]{fontenc}
\usepackage{newtxtext,newtxmath}
\usepackage{amsmath}
\usepackage{graphicx}
\usepackage{booktabs}
\usepackage{makecell}
\usepackage{tabularx}
\usepackage{subcaption}
\usepackage{xcolor}
\usepackage{appendix}
\usepackage{float}
\usepackage{enumitem}
\usepackage{cite}
\usepackage{caption}
\usepackage{url}
\usepackage[colorlinks=false,pdfborder={0 0 0}]{hyperref}
\usepackage{balance}

\providecommand{\orcidlink}[1]{}
\providecommand{\authormail}[1]{\href{mailto:#1}{#1}}
\makeatletter
\@ifundefined{remark}{%
  \newcounter{remarkctr}%
    {\par\medskip}%
  \@addtoreset{remarkctr}{section}%
}{}
\makeatother
\providecommand{\listheading}[1]{\noindent\textbf{#1}\par\smallskip}

\usepackage{cuted}
\usepackage{dblfloatfix}
\begin{document}

\twocolumn[{%
\begin{center}
{\large\bfseries Horizon-Dependent Tube MPC for Elliptical-Orbit Rendezvous Under Mass Uncertainty\par}
\vskip 0.93em
{\normalsize
\textbf{Omer Burak Iskender}\textsuperscript{a,*},\ \textbf{Keck-Voon Ling}\textsuperscript{a}\par}
\vskip 0.93em
{\normalsize\textsuperscript{a}\,\textit{School of Electrical and Electronic Engineering, Nanyang Technological University, Singapore. {E-mail:~\authormail{iske0001@e.ntu.edu.sg},~\authormail{ekvling@ntu.edu.sg}}}\par}
\vskip 0.31em
{\normalsize\textsuperscript{*}\,Corresponding author\par}
\vskip 0.62em
\noindent\rule{\linewidth}{0.4pt}
\vskip 0.25em
{\small\itshape \textbf{Accepted for publication.} 77th International Astronautical Congress (IAC 2026), Antalya, T\"{u}rkiye, 5--9 October 2026; paper IAC-26,C1,4,9,x108305. This is the accepted version (the authors' preprint) and not the version of record.\par}
\vskip 0.25em
\noindent\rule{\linewidth}{0.4pt}
\end{center}
\vskip 0.62em
\begin{center}
\textbf{Abstract}
\end{center}
\noindent
A spacecraft closing on a target from three hundred kilometres to contact flies one guidance law across five orders of magnitude of range, and a controller that is provably safe at close range can lose that guarantee completely at long range while continuing to fly as though nothing were wrong. This paper derives the range at which the guarantee lapses and uses it as a design rule. The bound compares the prediction model's own linearisation error against the disturbance set the controller was built to reject, and needs only the sampling period, the orbit and that disturbance bound, so it can be evaluated before any simulation. On a Mars Sample Return approach it disqualifies the homing phase, where most of the propellant is spent, and clears the other two. Re-posing the disqualified phase in relative orbital elements restores the guarantee; re-posing a phase the rule already clears, in a frame two orders of magnitude more accurate, changes propellant by under a tenth of one per cent, and it is that second prediction that makes the rule falsifiable rather than descriptive. The constraint tightening also ties the prediction horizon to feasibility, so the horizon search limit becomes a mission parameter rather than a solver setting. Against a reimplementation of a published benchmark that reproduces its propellant to within one per cent, over five hundred dispersed Monte Carlo transfers per case on matched seeds, the controller saves $29\%$ of the propellant on a circular target orbit and $40\%$ on an eccentric one, docking inside the $0.20$~m capture requirement on essentially every draw at a median miss near $5$~cm. Two results run the other way: the saving is bought with time of flight and computation, and it comes from what the guarantee demanded of the terminal condition rather than from better disturbance rejection. Recursive feasibility and asymptotic stability are not claimed.
\par
\vskip 0.93ex
\noindent\textbf{Keywords:} tube MPC, Yamanaka--Ankersen, eccentric orbit, mass uncertainty, spectral radius, phased rendezvous, robust guidance
\vskip 1.55ex
}]
\begin{strip}
\noindent
\begin{minipage}[t]{0.47\textwidth}
\listheading{Nomenclature}
\noindent\small
$A_k,B_k$ \quad YA discrete state-space matrices \\
$A_{\mathrm{cl},k}$ \quad $A_k+B_kK$, closed loop \\
$\bar A$ \quad $\max_k |A_{\mathrm{cl},k}|$, element-wise \\
$\bar{\mathbf e}_j$ \quad horizon-dependent error bound \\
$\bar{\mathbf w}_k$ \quad per-step disturbance bound \\
$\delta m_{\max}$ \quad fractional mass uncertainty bound \\
$K$ \quad ancillary feedback gain \\
$Q,R$ \quad MPC stage cost weights \\
$P,\alpha$ \quad terminal cost and terminal-set scaling \\
$\rho(\cdot)$ \quad spectral radius \\
$e,\nu$ \quad orbital eccentricity, true anomaly \\
$e^\star$ \quad certified eccentricity threshold \\
$T_s,N$ \quad sampling period, prediction horizon \\
$s^\star$ \quad predictor validity limit on separation
\normalsize
\end{minipage}\hfill
\begin{minipage}[t]{0.47\textwidth}
\listheading{Acronyms/Abbreviations}
\noindent\small
ADR: active debris removal \\
CW: Clohessy--Wiltshire \\
DARE: discrete algebraic Riccati equation \\
GTO: geostationary transfer orbit \\
LOS: line of sight \\
LP: linear program \\
LVLH: local vertical local horizontal \\
MC: Monte Carlo \\
MIB: minimum impulse bit \\
MPC: model predictive control \\
MSR: Mars Sample Return \\
QP: quadratic program \\
RPI: robust positively invariant \\
TH: Tschauner--Hempel \\
YA: Yamanaka--Ankersen
\normalsize
\end{minipage}
\end{strip}

\section{Introduction}

Rendezvous on an eccentric orbit combines two robustness challenges that the standard tube MPC formulations were not built to handle jointly. The prediction dynamics are linear time-varying: the Clohessy--Wiltshire equations are inadmissible because true anomaly does not advance uniformly, and the Yamanaka--Ankersen (YA) state transition matrix \cite{Yamanaka2002,Tschauner1965} is the natural replacement. The input matrix $B_k$ also depends on spacecraft mass, which drops through propellant depletion (typically $5$--$10\%$ over a rendezvous campaign) and steps through payload exchange in ADR or OOS missions \cite{Biesbroek2021,Forshaw2019}. The combination breaks the constant-matrix assumptions behind rigid-tube MPC for spacecraft.

Hartley et~al.~\cite{Hartley2012} set the template for an implementable MPC rendezvous system by partitioning the manoeuvre into a small number of guidance phases, each with its own sampling period, horizon, and constraint corridor. We adopt that phase architecture for the Mars Sample Return elliptical orbit and replace the per-phase nominal MPC with a horizon-dependent tube MPC, so the same guidance schedule now carries a robustness certificate rather than a fixed back-off. The figures in this paper deliberately mirror the diagnostics a rendezvous study is read for (phase-by-phase fuel, trajectory and capture plots, and sensitivity sweeps), computed here for the eccentric-orbit tube formulation.

\subsection*{Tube MPC foundations and the LTV gap}

Tube-based MPC originated as a robust extension of nominal MPC for constrained linear time-invariant systems \cite{Langson2004,Mayne2005}. The construction splits the controlled state into a nominal trajectory tracked by an inner-loop MPC and an error bounded by a robust positively invariant set: classically a minimal RPI (mRPI) approximation \cite{Rakovic2005}, computed once and held fixed across the horizon. Both the general survey of Mayne \cite{Mayne2014} and the aerospace-focused one of Eren et~al.~\cite{Eren2017} treat this LTI construction; neither addresses the time-varying closed loop, which is where a single worst-case width becomes wasteful. Step-by-step tightening predates the tube formulation itself, in the restricted-constraint MPC of Chisci et~al.~\cite{Chisci2001}, and was carried to generic LTV and uncertain nonlinear systems by Bumroongsri \cite{Bumroongsri2015} and K\"ohler et~al.~\cite{Kohler2021}, in each case replacing the single mRPI evaluation with a propagated sequence. Constraint tightening has been applied to rendezvous before \cite{Iskender2018Constraints}, though on circular-orbit dynamics and without a convergence certificate for the tightening itself. This paper specialises that machinery to the YA model and folds mass uncertainty into the same recursion.

\subsection*{Robust MPC for spacecraft proximity operations}

Closed-loop MPC for rendezvous and docking has been demonstrated at LEO \cite{Weiss2015}, in elliptical orbits \cite{Hartley2015,Hartley2012}, on air-bearing testbeds \cite{Mammarella2018,Buckner2018}, and in dual-quaternion form for coupled translation--attitude docking with an uncooperative target \cite{Iskender2019,Iskender2020}. Its spacecraft record is not confined to proximity operations: predictive detumbling of CubeSats has been benchmarked against the classical B-dot law under time-varying magnetic torque authority \cite{IskenderDetumbling2026}. Tube-based variants for spacecraft have used constant polytopic tubes \cite{Specht2023,Mammarella2018}, tubes resized online from an identified disturbance bound \cite{Oestreich2023}, and sector-bounded-nonlinearity formulations \cite{Bokor2025}. Each retains either an LTI assumption or a worst-case constant tube width, leaving the orbit-phase-specific tightening of the YA dynamics unexploited. The companion journal paper \cite{IskenderAISR2026} establishes the formal theory on this same YA model (recursive feasibility, ellipsoidal terminal set, ISS); augmented-dynamics extensions to fuel slosh \cite{IskenderActa2026,Iskender2026Sloshing} and structural flexibility \cite{IskenderJGCD2026} are developed separately.

\subsection*{Contribution}

What the paper proposes is a horizon-dependent tube MPC for the phased rendezvous of Ref.~\cite{Hartley2012}: the constraint tightening grows with the prediction step instead of being fixed at a worst-case constant, so the horizon the guidance selects and the robustness margin it is left with are decided inside one enumeration rather than in sequence. Everything around it is the benchmark's and stays that way. The phase schedule, the sampling periods, the horizon objective and its $w_u$ are adopted unchanged, so what the comparison isolates is the tightening and the cost form it makes affordable. Three results follow, and the first is what makes the other two more than bookkeeping. A box-tightened tube is certifiable only while the predictor's own linearisation residual stays inside the disturbance bound the tube was built against, which for the Yamanaka--Ankersen model bounds separation at $\sqrt{\bar w a/C}/(nT_s)$: $26.4$~km at a $300$~s period, and $13.2$~km at the $600$~s period the homing phase is actually sampled at. That phase opens at $300$~km, $23$ times outside its own limit, and its worst one-step residual overruns $\bar w$ by factors of $497$ and $852$; moving it to element space restores the certificate with $15$ to $28\%$ margin. The bound is what makes this predictive rather than anecdotal: it says in advance which reformulations can pay, and both directions were measured. Re-posing the closing phase in a cylindrical frame that is $128$ times more accurate buys $0.09\%$ of propellant, because that phase already sits inside the bound.

Second, the horizon and the robustness margin are chosen together. The variable horizon is not ours: the benchmark's guidance already minimises $J=N+w_u\|\mathbf u\|_1$ over $N\le N_{\max}$, and we take that functional and its $w_u$ unchanged (Section~\ref{sec:horizon}). What the tube changes is the set the search runs against, because the terminal tolerance left at horizon $N$ is $\mathrm{tol}_j-\bar e_j(N)$ and shrinks as $N$ grows, so reach and margin are traded inside one enumeration instead of in sequence. The consequence is measurable and was found by chasing a single stalled draw rather than by tuning: at a cap of $25$ steps, $11$ of $21$ stress cells flew the approach and one Monte Carlo draw in twenty-five stopped $908$~km out having spent nothing; at $50$, all $21$ flew, and the longer cap was $15.8$~m/s cheaper on the eleven cells both flew. The failure boundary is a joint condition on semi-major-axis error and target true anomaly, which is why no single-variable sweep located it.

Third, the resulting controller closes the approach for $24.1$ and $31.4$~kg less than the best replication of the benchmark, paired seed by seed, while arriving inside the requirement more often rather than less. Two tests have to be kept apart here and the earlier drafts of this paper welded them together. The free drift after blinding either crosses the target plane or it does not, and a crossing either falls inside the $0.20$~m requirement or it does not; a draw can do the first and fail the second. Over $500$ dispersed transfers per scenario at full fidelity the proposed form arrives inside the requirement on $499$ of $500$ circular draws and $500$ of $500$ elliptical ones, where the three replication variants manage $468$, $463$ and $452$ of $500$ on the elliptical scenario, intervals that do not reach $100\%$ (Section~\ref{sec:results}). The tail splits the same way and the paper claims no more than that: the $95$th-percentile miss is $10.39$~cm elliptical against $32.55$, a factor of $3.1$, but $10.73$~cm circular against $10.18$. Solving translation and pointing in one program is what buys the tail: on a $25$-draw paired ablation, dropping the coupling costs $0.02$~kg and opens the same percentile from $8.70$ and $7.94$~cm to $10.87$ and $12.18$.

The eccentricity certificate that earlier work reported as a scalar is the fourth result, and it is negative. It is insensitive to where the ancillary gain is designed and to the design weights, but it runs from $0.851$ at $T_s\!=\!50$~s to $0.249$ at $900$~s (Section~\ref{sec:cert}), so it must be quoted with the sampling period attached. Full formulations for every controller, plant and dispersion term summarised here are collected in a companion reference \cite{IskenderAISR2026}, transcribed from the implementation rather than from this manuscript.

\section{Relative Dynamics on an Eccentric Orbit}\label{sec:dyn}

\subsection*{Frames and the YA model}

\begin{figure*}[tbp]
\centering
\includegraphics[width=0.946\textwidth]{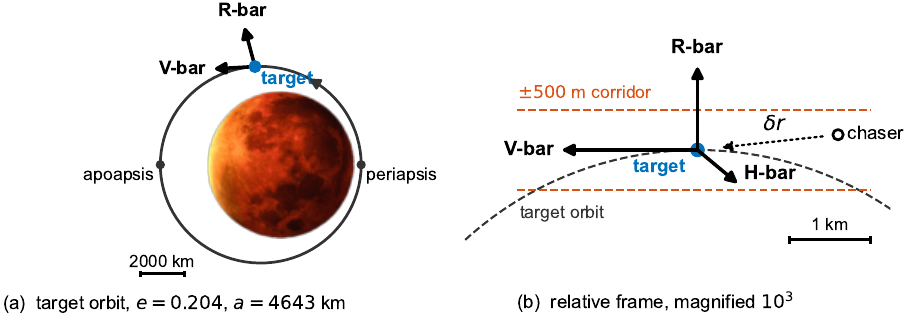}
\caption{The target orbit and the frame it carries. (a) The MSRE ellipse to scale against Mars. (b) The same geometry three orders of magnitude closer.}
\label{fig:lvlh}
\end{figure*}

The chaser state is expressed in the target Local-Vertical Local-Horizontal (LVLH) frame, with V-bar along-track, H-bar cross-track, and R-bar radial (Fig.~\ref{fig:lvlh}). Because the target follows a Keplerian ellipse, the natural independent variable is the true anomaly $\nu$, which advances per Kepler's second law $\dot\nu(\nu)=n(1+e\cos\nu)^2/(1-e^2)^{3/2}$ with mean motion $n=\sqrt{\mu/a^3}$. Let $\mathbf x_k\in\mathbb R^6$ denote the relative state at $t_k$. Under the YA state transition matrix at sampling period $T_s$ \cite{Yamanaka2002}, the prediction model is
\begin{equation}\label{eq:dyn}
\mathbf x_{k+1}=A_k\mathbf x_k+B_k\mathbf u_k+\mathbf w_k,
\end{equation}
where $\mathbf u_k\in\mathbb R^3$ is the impulsive $\Delta v$ applied at the start of the interval (hence $B_k=A_k\,[\mathbf 0_3;\,I_3]$, with $[\mathbf 0_3;\,I_3]\in\mathbb R^{6\times3}$ the stacked zero and identity blocks, depends on $k$ through $A_k$) and $|\mathbf w_k|\le\bar{\mathbf w}$ element-wise. The matrices $A_k$ depend on $\nu_k$ and are not constant.

Two features of Fig.~\ref{fig:lvlh} carry into everything that follows. The radial and along-track directions are separated from the local horizontal by the flight path angle, which reaches $11.7^\circ$ at $\nu=105^\circ$ on this orbit, so a frame built on the position vector and one built on the velocity vector do not coincide and the distinction has to be carried in the model. And the orbit arc drawn in panel~(b) is schematic: across the width of that panel the true orbit sags by $0.72$~m, three hundredths of one per cent of the panel height, so at close range the target's path really is a straight line to the accuracy anything here cares about. That smallness is what lets a linearised relative model work at all, and its failure at long range is the subject of Section~\ref{sec:valid}.

The cost of ignoring eccentricity is direct, and it is what forces the YA matrices rather than a preference for them. Propagating a free-drift relative state under a nonlinear two-body integration, under the YA matrices and under the constant-coefficient CW approximation on one semi-major axis separates the two approximations only when $e>0$. On a circular orbit the YA and CW errors against the nonlinear truth are identical at about $7$~m, second order in the $0.5$~km offset. On the MSRE orbit YA tracks the truth to $30$~m over one orbit while CW diverges to $8.6$~km, a factor of $280$, so at this eccentricity CW is inadmissible as a prediction model whatever controller is wrapped around it.

\subsection*{Mass-uncertain input matrix}

Mass changes through propellant depletion and payload exchange. We model the resulting input-matrix uncertainty as a multiplicative scalar,
\begin{equation}\label{eq:mass}
B_k^{\mathrm{true}}=(1+\delta_m)B_k,\qquad |\delta_m|\le\delta m_{\max},
\end{equation}
with $\delta m_{\max}\in[0.02,0.10]$ across the disturbance tiers. The uncertainty is persistent within a manoeuvre (a calibration error, not per-step noise) and unknown to the controller. One disclosure belongs with the model rather than with a limitations list at the end. The frozen $500$-draw campaign of Section~\ref{sec:results} was flown with this lifting switched off, \texttt{mass\_err\_max}~$=0$ at the sole construction site, with no cost specification overriding it and no mass channel at all in the element-space homing controller, so every tightening reported below is the additive bound $\bar{\mathbf w}$ alone and \eqref{eq:mass} is a design-time statement this campaign does not exercise. Runs with the term active are the obvious next campaign and are not claimed as completed.

\subsection*{Constraints and tube decomposition}

State and input satisfy element-wise box constraints,
\begin{equation}\label{eq:con}
\underline{\mathbf x}\le\mathbf x_k\le\bar{\mathbf x},\qquad |\mathbf u_k|\le\mathbf u_{\max},
\end{equation}
specifying the corridor half-width and per-axis $\Delta v$ envelope. Tube MPC decomposes the state as $\mathbf x_k=\mathbf z_k+\mathbf e_k$: the nominal trajectory $\mathbf z_k$ obeys the disturbance-free recursion, while the error $\mathbf e_k$ collects the closed-loop response to $\mathbf w_k$ and $\delta_m$ under the ancillary law $\mathbf u_k=\mathbf v_k+K\mathbf e_k$. The nominal trajectory is reset to the measurement at every step, $\mathbf z_0=\mathbf x_k$, so $\mathbf e_0=\mathbf 0$ and the applied control is $\mathbf u_k=\mathbf v_0^\star$ exactly: the ancillary term vanishes at the step that is executed. What $K$ determines is $A_{\mathrm{cl},k}=A_k+B_kK$, and through it the tightening recursion and the input back-off $|K|\bar{\mathbf e}_j$. The error dynamics are
\begin{equation}\label{eq:err}
\mathbf e_{k+1}=A_{\mathrm{cl},k}\mathbf e_k+\mathbf w_k^{\mathrm{eff}},\quad
\mathbf w_k^{\mathrm{eff}}=\mathbf w_k+\delta_m B_k\mathbf u_k,
\end{equation}
with $A_{\mathrm{cl},k}=A_k+B_kK$. The second term is the multiplicative-to-additive lifting: it bounds the input-matrix uncertainty by a per-step additive contribution scaled by $|B_k|$ and the actuator envelope.

\section{Phased Rendezvous Mission Design}\label{sec:phased}

\begin{figure*}[tbp]
\centering
\includegraphics[width=0.78\textwidth]{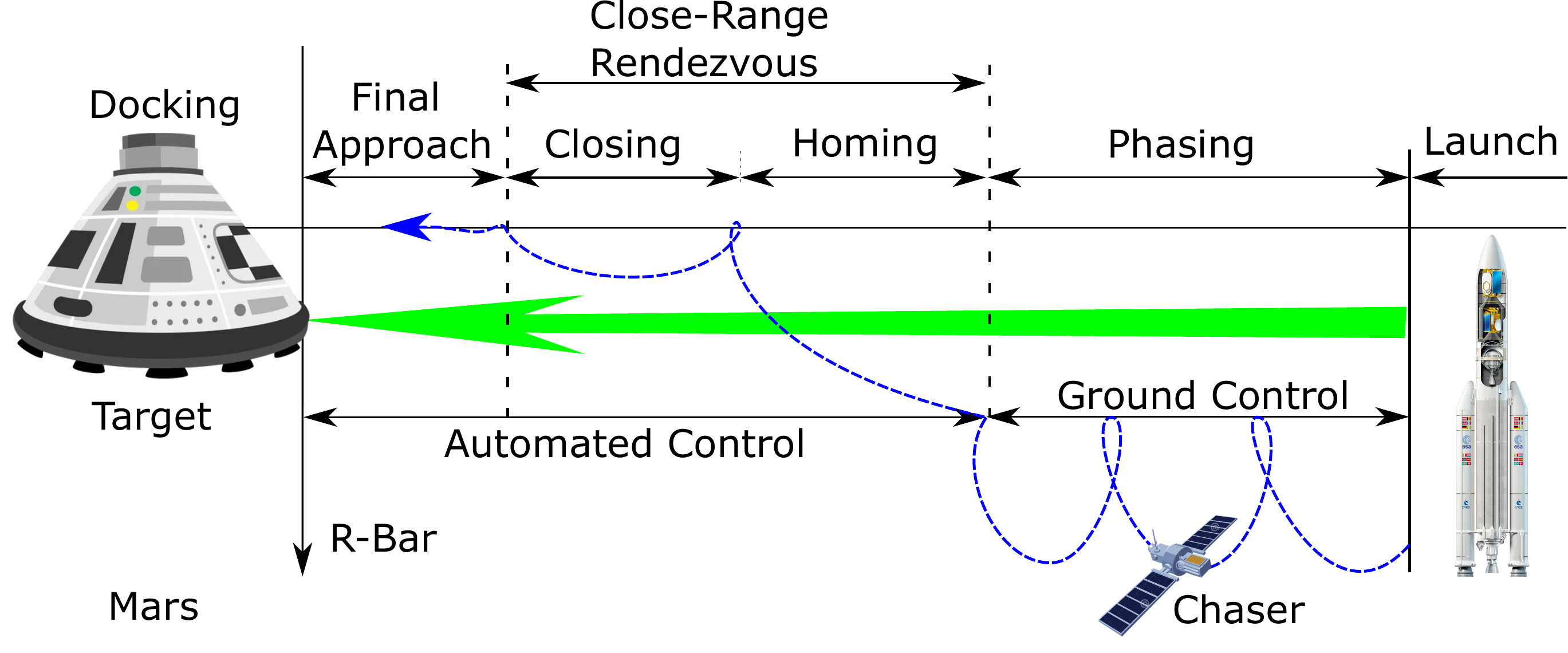}\\[4pt]
{\small (a)}\\[6pt]
\includegraphics[width=0.86\textwidth]{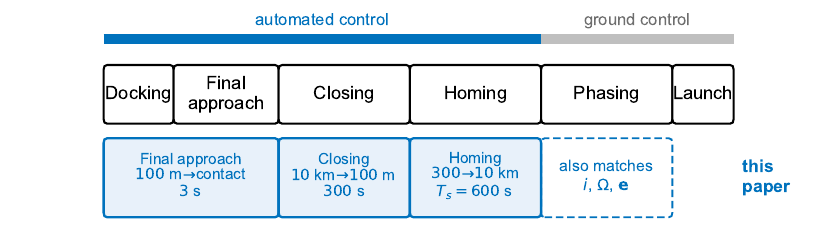}\\[2pt]
{\small (b)}
\caption{The standard rendezvous sequence and the three guidance phases used here.}
\label{fig:process}
\end{figure*}

Phasing precedes the scenario studied here: two phasing orbits consume the phase angle and the chaser transfers to the target orbit, after which far-range rendezvous begins. The campaign of Section~\ref{sec:results} starts at that point, with the chaser already near the target orbit and $300$~km behind it, so everything earlier is context rather than part of the problem posed here. The approach runs on the Hartley scenario \cite{Hartley2012} as specified there, rather than on a ladder of our own, so that every controller reported here can be measured against a published baseline on identical geometry. Three guidance phases divide it, each run by the tube MPC of Section~\ref{sec:cert}, and Fig.~\ref{fig:process} places them inside the standard rendezvous sequence \cite{Fehse2003,Iskender2020}. \emph{Homing} closes the bulk of the along-track gap from $300$~km at a $600$~s period; \emph{closing} takes the hand-over near $10$~km down to a $100$~m target approach point at $300$~s; the \emph{final approach} completes the run to contact at a $3$~s period. One correspondence is inexact and it is the interesting one: homing in the standard sequence closes range under relative navigation, whereas the homing phase here additionally drives inclination, right ascension and the eccentricity vector to zero, which that sequence assigns to phasing. That straddle is why this phase, and only this phase, is re-posed in relative elements in Section~\ref{sec:valid}. Only the final approach carries a fixed horizon. The two long-range phases re-select $N$ at every guidance call by enumerating $[n_{\min},n_{\max}]=[8,50]$ and scoring each candidate, which is the benchmark's own construction \cite{Hartley2012} rather than a departure from it; Section~\ref{sec:horizon} states what the tightening adds to that rule and what it costs. Table~\ref{tab:phases} gives the schedule together with the admissibility each phase inherits from its own sampling period, and those last two columns are the reason the schedule cannot be separated from the certificate.

Reading Table~\ref{tab:phases} down the $e^\star$ column is the clearest argument against quoting the threshold as one number. The same vehicle on the same orbit is certified with a margin of $1.77$ at the homing period and with essentially unlimited margin at the terminal period, because $e^\star$ falls as $T_s$ rises. MSRE clears every row, but an orbit at $e=0.5$ would be certified by a headline figure of $0.66$ and would fail outright at the $600$~s period, where the true threshold is $0.362$. The sampling period is therefore not a free implementation choice; it is part of what is being certified.


\begin{table}[tbp]
\centering
\caption{Phased schedule, with the admissibility each phase inherits from its own sampling period. $e^\star$ is the threshold of \eqref{eq:bis} at that $T_s$.}
\label{tab:phases}
\setlength{\tabcolsep}{3pt}
\resizebox{\columnwidth}{!}{%
\begin{tabular}{@{}lccccccc@{}}
\toprule
Phase & Range & $T_s$ & $N$ & $u_{\max}$ & $e^\star$ & $\rho$ & marg. \\
 & & [s] & & [m/s] & $(T_s)$ & & $e^\star\!/e$ \\
\midrule
Homing  & 300$\to$30--10\,km & 600 & 8--50 & 5.3875 & 0.362 & 0.472 & 1.77 \\
Closing & 30--10\,km$\to$100\,m & 300 & 8--50 & 2.6937 & 0.552 & 0.205 & 2.70 \\
Final   & 100\,m$\to$cont. & 3 & 15 & 0.0269 & 0.972 & 0.012 & 4.75 \\
\bottomrule
\end{tabular}}
\end{table}


The three phases do not partition range, and reading Table~\ref{tab:phases} as though they did produces an apparent $30$ to $10$~km gap that does not exist. Hand-over from homing is a conjunction: the in-track separation must lie in $[10,30]$~km \emph{and} the orbit-match residual of the five non-along-track element rows, expressed as an equivalent length, must fall below $R_h=2000$~m. The band is where the two conditions become satisfiable at once. What the conjunction protects is the synchronisation the homing terminal rows bought, since the Yamanaka--Ankersen model that closing predicts with is a local relative description and cannot recover an inclination error at all. The remaining hand-overs are pure range tests: at $\rho=100$~m the guidance switches model and sampling period together, and at $\rho=3$~m control stops and the vehicle coasts to the interface, so the last gate marks the end of actuation rather than a change of controller. That coast is why every capture number in Section~\ref{sec:results} is scored on a free-drift crossing of the target plane and not on a commanded state.

A rotating keep-out constraint is not part of the Mars approach as flown, and saying so plainly is worth more than a figure of one that is not enforced. The state box the campaign carries is inert at $\pm10^9$~m, so the tube acts through the terminal element tolerance \eqref{eq:termrow} and the input box alone. Nothing in Section~\ref{sec:results} should be read as a corridor being consumed, and the corridor diagnostic the tightening is more usually sold on, together with the cross-body envelope and the geostationary co-location study, is carried in the companion journal treatment \cite{IskenderAISR2026} rather than compressed into this paper.


\section{Horizon-Dependent Tube MPC and the Eccentricity Certificate}\label{sec:cert}

\subsection*{Tightened-constraint QP}

At each instant $k$ the controller solves
\begin{equation}\label{eq:qp}
\begin{aligned}
\min_{\mathbf z, \mathbf v}\;& \!\sum_{j=0}^{N-1}\!\left(\|\mathbf z_j-\mathbf z_j^{\mathrm{ref}}\|_Q^2+\|\mathbf v_j\|_R^2\right)+\|\mathbf z_N-\mathbf z_N^{\mathrm{ref}}\|_P^2 \\
\text{s.t.}\;& \mathbf z_{j+1}=A_{k+j}\mathbf z_j+B_{k+j}\mathbf v_j,\;\mathbf z_0=\mathbf x_k,\\
& \underline{\mathbf x}+\bar{\mathbf e}_j\le\mathbf z_j\le\bar{\mathbf x}-\bar{\mathbf e}_j,\\
& |\mathbf v_j|\le\mathbf u_{\max}-|K|\bar{\mathbf e}_j,
\end{aligned}
\end{equation}
where $\mathbf z_j^{\mathrm{ref}}$ is the per-phase glide-slope reference, $Q\succeq 0$, $R\succ 0$, and $P$ is the terminal cost from the DARE. Every constraint in \eqref{eq:qp} is linear, so the problem is a quadratic program and OSQP \cite{Stellato2020} solves it directly; the condensed form carries $3N$ variables. The horizon-dependent tightening propagates one step at a time,
\begin{equation}\label{eq:emax}
\bar{\mathbf e}_{j+1}=|A_{\mathrm{cl},k+j}|\,\bar{\mathbf e}_j+\bar{\mathbf w}_j,\quad\bar{\mathbf e}_0=\mathbf 0,
\end{equation}
with the per-step disturbance bound
\begin{equation}\label{eq:wj}
\bar{\mathbf w}_j=\bar{\mathbf w}+\delta m_{\max}|B_{k+j}|\,\mathbf u_{\max}.
\end{equation}
Additive noise and mass uncertainty enter \eqref{eq:emax}--\eqref{eq:wj} as a single box, propagated step by step with element-wise arithmetic and no set computation. Which of the two dominates is a scheduling question rather than a modelling one. The mass term carries the sampling period twice, once through $B_k=A_k[\mathbf 0_3;I_3]$ and once through $u_{\max}=T_sF_{\max}/m$, so it grows as $T_s^2$: at the homing period it is $2.8$ times the additive bound and at the FINAL period only a third of it, which is a property of \eqref{eq:wj} as designed rather than of the campaign, since the runs reported here carry the mass term switched off as Section~\ref{sec:dyn} records. What follows for \eqref{eq:qp} is robust constraint satisfaction, and nothing beyond it. If the tightened program is feasible and $|\mathbf e_j|\le\bar{\mathbf e}_j$ element-wise for every $j\le N$, then $\mathbf x_j=\mathbf z_j+\mathbf e_j$ meets the original box row by row, and so does $\mathbf u_j$, by the triangle inequality applied to each row \cite[Theorem~1]{IskenderAISR2026}. The cost appears nowhere in that argument, which is why the guarantee is indifferent to which objective form Section~\ref{sec:results} compares. Asymptotic stability is not claimed here: $P$ solves the DARE and supplies the unconstrained infinite-horizon cost, but with no terminal set it is not a Lyapunov function for the constrained problem.

Recursive feasibility is a separate matter, and \eqref{eq:qp} does not deliver it. The proof in \cite[Theorem~2]{IskenderAISR2026} needs a terminal ingredient: an ellipsoidal robust positively invariant set $\mathcal X_f=\{\mathbf x:(\mathbf x-\mathbf x_s)^\top P(\mathbf x-\mathbf x_s)\le\alpha\}$ built offline at the worst-case orbit phase, enforced as
\begin{equation}\label{eq:alphaN}
\begin{aligned}
&(\mathbf z_N-\mathbf z_N^{\mathrm{ref}})^\top P(\mathbf z_N-\mathbf z_N^{\mathrm{ref}})\le\alpha_N,\\
&\alpha_N=\left(\sqrt\alpha-\big\|\,|P^{1/2}|\,\bar{\mathbf e}_N\big\|_2\right)^2,
\end{aligned}
\end{equation}
where the level is deflated by the tube radius at $j\!=\!N$ so that the true terminal state, not merely the nominal one, lands in $\mathcal X_f$; the element-wise $|P^{1/2}|$ keeps that bound valid for every error inside the box $\bar{\mathbf e}_N$. What that costs is easy to state and hard to pay. Sizing $\alpha$ as the largest level whose ellipsoid still fits the final-approach state box gives $\sqrt\alpha=2.82$, and a $\bar w_{\mathrm{pos}}=20$~m, $\delta m_{\max}=5\%$ design deflates it by $2.06$, so $73\%$ of the terminal radius is spent on the tube before the nominal is allowed anywhere. Adding \eqref{eq:alphaN} also changes the problem class: the constraint is quadratic, so \eqref{eq:qp} becomes a quadratically constrained program rather than a QP, and OSQP cannot accept it. Replacing the ellipsoid by the axis-aligned box $|(\mathbf z_N-\mathbf z_N^{\mathrm{ref}})_i|\le\sqrt{\alpha_N/P_{ii}}$ keeps the problem a QP but circumscribes the ellipsoid, which relaxes the constraint instead of enforcing it and does not transfer the theorem. A solver with native second-order-cone support, or an inscribed polytope, would be needed for that. The campaigns reported here run \eqref{eq:qp} as printed, without any terminal constraint, so recursive feasibility is a design result this paper does not exercise. The recursion differs from the constant mRPI width it converges to by less than one might expect, and the companion journal paper \cite{IskenderAISR2026} plots the two profiles against each other: because $\rho(|A_{\mathrm{cl},k}|)$ is small on this orbit, the recursion reaches the mRPI width within a single step, and from $j\!=\!1$ onward the two profiles agree to between $1$ and $16\%$. The whole of the difference sits at $j\!=\!0$, where $\bar{\mathbf e}_0=\mathbf 0$ leaves the corridor and the actuator envelope untightened. That step is the one that matters operationally, because the applied control is $\mathbf u_k=\mathbf v_0^\star$: the recursion hands the controller the full $\pm5$~m/s per axis at the executed step, against $3.36$~m/s for the constant-width tube.

\subsection*{Spectral stability certificate}

The error recursion \eqref{eq:emax} is guaranteed to stay bounded when the worst-case element-wise closed-loop matrix $\bar A=\max_k|A_{\mathrm{cl},k}|$ satisfies $\rho(\bar A)<1$. The condition is sufficient, not necessary: $\bar A$ majorises every $|A_{\mathrm{cl},k}|$ element-wise, so a time-varying sequence can converge even where $\rho(\bar A)\ge1$. It is also strictly stronger than $\rho(|A_{\mathrm{cl},k}|)<1$ holding at each $k$ separately, and must be checked in its own right; for the MSRE design the two are $0.129$ and $0.128$ \cite{IskenderAISR2026}. We design the ancillary gain $K$ at $\nu_d=0$ (periapsis), where the dynamics are most active so the gain pre-compensates the largest single-step magnitudes; the sensitivity to this choice is quantified below. With $Q_K=\mathrm{diag}(10^3,10^3,10^3,10,10,10)$, $R_K=I_3$ on the MSRE orbit, the periapsis-designed gain is near-deadbeat at its design point and least matched at apoapsis: $\rho(|A_{\mathrm{cl},k}|)$ is minimal at periapsis ($\approx2\!\times\!10^{-4}$) and peaks at apoapsis ($0.128$), a profile the companion journal paper \cite{IskenderAISR2026} plots over the orbit. Under the PF condition the steady-state tightening obeys $\bar{\mathbf e}_\infty\le(I-\bar A)^{-1}\bar{\mathbf w}_\infty$ element-wise, so the limit is finite and the tube exists; the right-hand side is an upper bound, not the realised width, because it substitutes the orbit-worst-case $\bar A$ for the actual $|A_{\mathrm{cl},k+j}|$ at every step.

\subsection*{Eccentricity threshold}

The threshold is recovered by bisecting $\rho(\bar A(e))=1$ over $e\in[0,1)$:
\begin{equation}\label{eq:bis}
e^\star=\sup\{e:\rho(\bar A(e))<1\},
\end{equation}
which evaluates to $e^\star=0.6455$ for the MSRE configuration at a $200$~s sampling period. The qualifier is not decoration. Bisecting \eqref{eq:bis} at each period gives $0.8514$ at $50$~s, $0.7686$ at $100$~s, $0.6455$ at $200$~s, $0.5517$ at $300$~s, $0.3617$ at $600$~s and $0.2492$ at $900$~s, so the threshold falls by a factor of $3.4$ across the range a phased approach schedules, and a single reported number is admissible only alongside the period it was computed at. Two conditions must also be kept apart: $\rho(\bar A)<1$, which is what \eqref{eq:emax} needs, returns $0.6455$, whereas maximising $\rho(|A_{\mathrm{cl},k}|)$ over anomaly separately returns $0.6609$ and is the weaker test.
Figure~\ref{fig:eccthreshold} plots the threshold against the period. The underlying curve $\rho(\bar A;e)$ inherits the structural form $\rho\propto((1+e)/(1-e))^2$ from Kepler's second law applied to the YA matrices: the periapsis-to-apoapsis rate ratio is $1.004$ at LEO, $2.30$ at Mars, and $42$ at GTO, and the threshold is in effect the inverse of this asymmetry. Mars ($e=0.204$) clears the condition at every period tested, though the margin is not uniform: $\rho(\bar A)$ at that eccentricity runs from $0.029$ at $50$~s to $0.791$ at $900$~s, so a phase sampled slowly enough sits close to losing the certificate on an orbit the headline number calls comfortable. An orbit at $e=0.5$ is a sharper illustration, because it holds the certificate up to $368$~s and loses it above, which places it on the wrong side of the homing period and the right side of the closing one in the same mission.

\subsection*{Design-point sensitivity}

A natural concern is whether $e^\star$ is an artefact of designing $K$ at periapsis. Sweeping the design true anomaly $\nu_d$ over six points around the orbit and recovering $e^\star(\nu_d)$ by bisection at each gives an envelope of $[0.635,0.664]$, a span of $4.4\%$, with the periapsis design least conservative and the apoapsis design most. Replacing the state and input weights outright, from $\mathrm{diag}(10^3,10^3,10^3,10,10,10)$ with $R_K=I$ to $\mathrm{diag}(10^2,10^2,10^2,1,1,1)$ with $R_K=10I$, leaves the threshold unchanged to four decimals at every period tested, so the insensitivity extends past the design anomaly to the weights themselves. What $e^\star$ is not invariant to is the sampling period, which moves it by a factor of $3.4$ over the same span; the certificate is a property of the discretised design rather than of the orbit alone, and it must be quoted with the period attached.

\section{Where the Tube Is Valid, and What Follows}\label{sec:valid}

\begin{figure*}[t]
\centering
\begin{minipage}[t]{0.49\textwidth}\centering
\includegraphics[width=\linewidth]{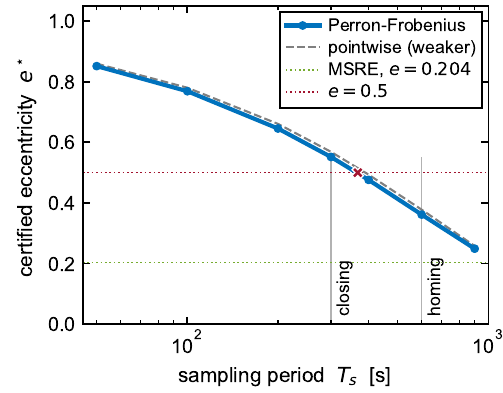}
\caption{$e^\star(T_s)$ under the Perron--Frobenius condition \eqref{eq:bis}, against the weaker pointwise condition earlier work reported.}
\label{fig:eccthreshold}
\end{minipage}\hfill
\begin{minipage}[t]{0.49\textwidth}\centering
\includegraphics[width=\linewidth]{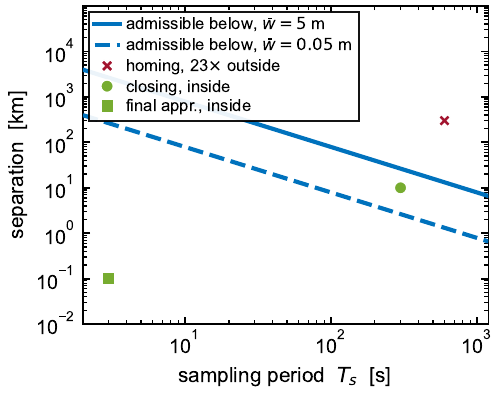}
\caption{The separation limit \eqref{eq:valid} against sampling period, each phase at its own period and opening range.}
\label{fig:valid}
\end{minipage}
\end{figure*}

The certificate of Section~\ref{sec:cert} bounds the tightening recursion, but it says nothing about whether the model the recursion is built on describes the vehicle. Box tightening treats every departure from the nominal prediction as a disturbance inside $\bar{\mathbf w}$, so it is sound only while the predictor's own linearisation residual stays inside that bound. On the YA model the residual over one step grows with the square of separation, $\varepsilon\approx0.866\,n^2T_s^2s^2/a$, and inverting it for $s$ gives the range over which the tube may be trusted,
\begin{equation}\label{eq:valid}
s\le\frac{1}{nT_s}\sqrt{\frac{\bar w\,a}{C}},
\end{equation}
which is $26.4$~km at the $300$~s closing period and, because the limit falls as $1/T_s$, only $13.2$~km at the $600$~s period homing runs at (Fig.~\ref{fig:valid}). Each phase must be measured against its own period, not against a headline one, and the $26$~km figure quoted in the abstract belongs to the closing phase. The bound is cheap to evaluate and it is the only statement here that constrains where the machinery of Section~\ref{sec:cert} may be applied at all.

Measured against \eqref{eq:valid}, the homing phase is not marginal, it is disqualified. It opens at $300$~km against a limit of $13.2$~km, a factor of $23$, and the residual that follows is worse than the ratio suggests because it grows quadratically: the Cartesian predictor errs by $2486$~m on the elliptical scenario and $4260$~m on the circular one against a $\bar w$ of $5$~m, so the worst one-step residual overruns its own bound by factors of $497$ and $852$. Earlier reports of this work quoted $4759$ and $6350$ for the same comparison; those figures are not the quotients of the residuals beside them and are withdrawn here rather than reconstructed. A certificate exceeded by two orders of magnitude is not conservative; it is void, and no amount of tuning inside the recursion recovers it.

 Re-posing that phase in relative orbital elements, with the terminal condition written as equalities on five elements rather than as a position box, brings the same ratio to $0.856$ and $0.716$, satisfied with $15$ to $28\%$ margin. The mechanism is the terminal condition rather than the change of coordinates: a position-only box at $20$~km is satisfied by the free drift itself, so the phase was burning $670$~kg to enforce a constraint that was never active.

The same bound predicts where such a reformulation cannot help, and that prediction was tested rather than assumed. The closing phase hands over near $10$~km and its worst one-step residual anywhere the controller actually flies is $1.46$~m circular and $1.65$~m elliptical, comfortably inside $\bar w$. Re-posing it in the cylindrical relative frame, which is $11$ to $128$ times more accurate over that range, changes propellant by $0.09\%$. The frame removes an error that never bound. Taken together the two experiments make \eqref{eq:valid} a design rule: reformulate the phases that violate it and leave the others alone, because inside the bound the tube cannot tell the difference.

What that rule is worth, and what ignoring it costs, are different sizes.
Table~\ref{tab:rule} prices both on the same mission. Keeping the Cartesian
predictor in the phase that violates the bound is not a small penalty paid for
a simpler formulation: the quadratic form spends $1185$~kg on the circular
scenario against $108$ in element space, a factor of eleven, and $264$ against
$92$ on the elliptical one, and it fails to reach the target approach point on
one draw in five where the element form reaches it on every draw. The $1$-norm
form behaves the same way, $1158$ against $69$~kg. A mission that ignores the
bound does not fly worse; on this scenario it does not fly.

The same rule, applied where the bound is satisfied, returns nothing, and that
is the half that makes it predictive rather than a slogan: the cylindrical
frame is $11$ to $128$ times more accurate over the closing phase and changes
propellant by $0.09\%$. Terminal accuracy is deliberately absent from
Table~\ref{tab:rule}. The Cartesian cells reach the interface on fewer draws,
so their miss statistics are taken over a different and easier subset, and at
five draws per cell the comparison would not be worth making even if the
subsets matched. Where accuracy is bought in this paper is the attitude
coupling of Section~\ref{sec:results}, not the certificate.

\begin{table}[tbp]
\centering
\caption{What the validity bound is worth. Five draws per cell, medians, at the
simulator of Table~\ref{tab:sim}. \emph{Reached} counts draws that arrived at
the target approach point.}
\label{tab:rule}
\small
\resizebox{\columnwidth}{!}{%
\begin{tabular}{@{}llcc@{}}
\toprule
Treatment & Form & kg & Reached \\
\midrule
\multicolumn{4}{@{}l}{\emph{Homing, $497$--$852\times$ outside the bound}} \\
\quad Cartesian, ignored & QP & $1185$ & $4/5$ \\
\quad Elements, followed & QP & $\mathbf{108}$ & $\mathbf{5/5}$ \\
\quad Cartesian, ignored & $1$-norm & $1158$ & $2/5$ \\
\quad Elements, followed & $1$-norm & $\mathbf{69}$ & $4/5$ \\
\midrule
\multicolumn{4}{@{}l}{\emph{Closing, inside the bound}} \\
\quad Cartesian, followed & QP & \multicolumn{2}{c}{reference} \\
\quad Cylindrical, wasted & QP & \multicolumn{2}{c}{$-0.09\%$} \\
\bottomrule
\end{tabular}}

\end{table}

\section{Horizon Selection and the Tightened Terminal Set}\label{sec:horizon}

The variable horizon is not a contribution of this work. The benchmark's guidance already solves, at every call,
\begin{equation}\label{eq:jn}
N^\star=\arg\!\!\!\min_{N\in[n_{\min},n_{\max}]}\!\!\!\! J(N),\;\; J(N)=N+w_u\|\mathbf u^\star(N)\|_1,
\end{equation}
so that each extra step of manoeuvre time costs one unit and has to repay it in propellant \cite[\S4.2]{Hartley2012}. We adopt \eqref{eq:jn} unchanged, including the benchmark's own $w_u$ rather than a retuned one, and apply the same rule to both objective forms compared below, so that the forms differ in what is minimised at fixed $N$ and never in how $N$ is chosen. One difference of implementation is worth recording: at each candidate $N$ the benchmark solves a linear program, the $1$-norm being its objective, whereas the inner problem here is the quadratic program \eqref{eq:qp} and \eqref{eq:jn} scores its solution.

What the tube changes is the set the search runs against. In the benchmark the constraint set seen at every candidate is the nominal one and does not depend on $N$, so a longer horizon is strictly more reachable and the only price is the $+1$. Under \eqref{eq:emax} the terminal row available at horizon $N$ is
\begin{equation}\label{eq:termrow}
|\delta e_N(j)|\le \mathrm{tol}_j-\bar e_j(N),
\end{equation}
whose right-hand side is eroded as $N$ grows, and the in-track acceptance window is closed from both sides by the tube's own position uncertainty: $2.5$~km of closure at $N\!=\!8$ against $20.3$~km at $N\!=\!25$, on a half-window of $10$~km. Feasibility therefore occupies an interval $N\in[\underline N(\delta\mathbf e),\overline N]$ rather than a half-line. Too short cannot reach, and too long is tightened out of existence. Horizon selection and robustness margin are solved together in \eqref{eq:jn}, which is the sense in which the tightening here is horizon-dependent; the horizon being variable is inherited, the horizon being an argument of the margin is not.

\subsection*{What the cap was silently deciding}

The lower end $\underline N$ rises with the dispersion, so a cap chosen for compute can remove a draw from the campaign altogether. One Monte Carlo draw in twenty-five did exactly that at $n_{\max}=25$: every candidate horizon was infeasible, the untightened fallback was infeasible as well, the solver returned nothing nine times in succession, and the run abandoned $908$~km out having spent no propellant. That is not a tuning failure and not an authority failure. From the Gauss variational equation a tangential impulse changes semi-major axis by $\delta a=2a^2v\Delta v_t/\mu$, which at $u_{\max}=3$~m/s removes up to $9.2$~km of $\delta a$ per burn on the elliptical scenario, so pure authority would need only six to eight steps. What binds is phasing: a semi-major-axis error is an in-track drift rate $\dot x_1=-\tfrac32 n\,\delta a$, and the largest error that can be flown into a $W=20$~km window within one sample is
\begin{equation}\label{eq:dastar}
\delta a^\star=\frac{2W}{3nT_s}=34~\text{km (elliptical)},\quad 26~\text{km (circular)}.
\end{equation}
Above $\delta a^\star$ a single step of drift overshoots the whole window, so the drift rate must first be bled down, and where the vehicle ends up while it is being bled depends on where it started. The failure boundary is a joint condition on $(\delta a,\delta\nu)$, not a threshold on either.

\begin{figure}[tbp]
\centering
\includegraphics[width=\columnwidth]{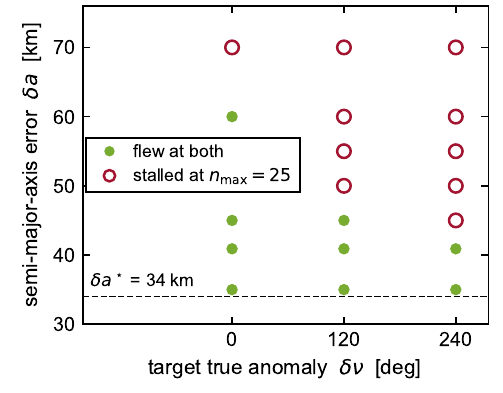}
\caption{The $21$ stress cells at full fidelity. Open markers spent nothing at $n_{\max}=25$ and flew at $50$.}
\label{fig:nmax}
\end{figure}

That prediction is what the stress grid of Fig.~\ref{fig:nmax} tests, on cells clustered where the controller broke rather than sampled at random, because a random draw puts almost no mass at the top of the dispersion range. Semi-major-axis errors from $35$ to $70$~km crossed with three target true anomalies give $21$ cells at full fidelity: $n_{\max}=25$ flew $11$ of them and $n_{\max}=50$ flew all $21$, with no cell made worse. A stalled cell is one that spent no propellant at all, having found no feasible horizon at any candidate, which is a different and more visible failure than arriving badly. The boundary is the staircase the argument above predicts, holding through $60$~km at $\delta\nu=0$ and breaking from $45$~km at $\delta\nu=240^\circ$, against the $34$~km of \eqref{eq:dastar}; the gap is the few steps of bleeding a longer horizon can afford. On the eleven cells both caps flew, the longer one is $15.8$~m/s cheaper paired and cheaper on all eleven, so the short cap was not only refusing hard problems but forcing worse plans on ones it could still solve. The only cost is $2.5$ times the wall clock (Table~\ref{tab:compute}).

The cap remains binding at $50$, and the selection history says so rather than the argument asserting it. Logged at every one of the $71\,398$ guidance calls of the $1000$ dispersed transfers of Section~\ref{sec:results}, the selected horizon has a median of $23$ steps but sits within one step of the cap on $22\%$ of them, and the pile-up is heavier on the circular scenario ($25\%$) than the elliptical one ($20\%$). The guidance is not choosing $50$; it is limited to $50$. Whether $75$ would help further has not been flown.

\section{Simulation Setup}\label{sec:setup}

\subsection*{How faithful the benchmark is}

Every comparison below is made against a reimplementation of Hartley's controller rather than against the published propellant, and the reason is reproducibility rather than disagreement. Reimplemented faithfully, with the reference $J_2$-modified GVE predictor and the initial hold applied before the first burn, the replication returns $102.27$~kg of homing and closing propellant against the reference $102.96$ on the circular scenario and $93.98$ against $89.29$ on the elliptical one, ratios of $0.993$ and $1.053$ (Table~\ref{tab:replication}). Measured instead against the digitised reference full-mission curve the same runs give $0.999$ and $1.056$, so the match does not depend on which of the two scopes is used, and that agreement is the licence for every comparison that follows. Both pieces of the reference controller have to be present for it. Dropping the initial hold alone takes the circular ratio from $0.993$ to $0.727$, and dropping the GVE predictor as well moves it back to $0.834$, which is a partial cancellation of two omissions rather than a recovery; an earlier draft measured the no-hold variant, read $0.82$, and concluded the replication had failed, when what it had done was leave out a piece of the reference controller.

\begin{table}[tbp]
\centering
\caption{The replication reproduces the published propellant on both scopes. One undispersed transfer from the Table~1 elements of Ref.~\cite{Hartley2012}.}
\label{tab:replication}
\setlength{\tabcolsep}{4pt}
\resizebox{\columnwidth}{!}{
\begin{tabular}{@{}llcccc@{}}
\toprule
Form & Scen. & Hom.$+$Clos. & vs Ref. & full & vs Ref. \\
 & & [kg] & text & [kg] & curve \\
\midrule
HARVD~\cite{Hartley2012} & C & -- & -- & 136.53 & 1.301 \\
Reference MPC~\cite{Hartley2012} & C & 102.96 & 1.000 & 104.92 & 1.000 \\
\textbf{Reference impl.} & C & \textbf{102.27} & \textbf{0.993} & \textbf{104.83} & \textbf{0.999} \\
Reference impl. (Kep.) & C & 85.92 & 0.834 & 88.63 & 0.845 \\
Reference impl. (GVE) & C & 74.85 & 0.727 & 77.69 & 0.740 \\
\textbf{Proposed} & C & 76.04 & 0.739 & 77.58 & 0.739 \\
\midrule
HARVD~\cite{Hartley2012} & E & -- & -- & 130.53 & 1.427 \\
Reference MPC~\cite{Hartley2012} & E & 89.29 & 1.000 & 91.47 & 1.000 \\
\textbf{Reference impl.} & E & \textbf{93.98} & \textbf{1.053} & \textbf{96.55} & \textbf{1.056} \\
Reference impl. (Kep.) & E & 62.09 & 0.695 & 64.94 & 0.710 \\
Reference impl. (GVE) & E & 59.40 & 0.665 & 61.90 & 0.677 \\
\textbf{Proposed} & E & 55.61 & 0.623 & 57.69 & 0.631 \\
\bottomrule
\end{tabular}
}
\end{table}

Quoting the reimplementation as the ruler, rather than the published propellant, keeps the plant, the frames, the phase ladder, the thruster model and the assumed specific impulse identical on both sides of every comparison, which is the only arrangement under which a difference can be attributed to the controller. It also means the comparison inherits whatever the reference controller does well: the replication is not a straw baseline chosen to be beaten, it is a reconstruction that lands on the published value and is then flown against ours on the same seeds.

The shortfall is understood well enough to be stated rather than merely disclosed, and it is not a defect of the objective. The program class matches the reference class in every phase: a linear program for the two long-range phases and a quadratic program for the last hundred metres, switching once at the target approach point, exactly as specified in \cite[\S8]{Hartley2012}. Nor is it a weighting. Sweeping the fuel weight across six decades saturates above $w_u\!\approx\!300$, and the saturated limit is a greedy minimum-fuel policy that still costs $85.7$~kg against the published $102.96$: a weight can move a solution along its own Pareto front but cannot push it above its own fuel optimum. What remains is the constraint set. The reference passive-safety half-spaces, the reference drift condition and the reference one-sample computational delay are all absent from the reimplementation because the safety radius $R_S(t)$ that parameterises them is never published, and every one of them can only raise the optimum of the program actually solved. The gap therefore has the sign the omissions predict, on both scenarios and in both long-range phases.

One consequence is visible in the trajectories and is worth stating because it looks like an error and is not. The published tracks spend $86\%$ of their length at positive in-track and reach $+154$~km, while the reimplementation approaches monotonically and never crosses the target. The initial condition carries a large natural drift, closing $296$~km unaided within two thirds of an orbit, and the reference first burn falls where that drift delivers the chaser rather than where a schedule places it. A hold that releases when coasting stops closing the range faster than the thrusters could reproduces both the reference epoch and the reference geometry, at a cost of $19\%$ more propellant on the circular scenario and $51\%$ on the elliptical one. That the faithful variant is the expensive one is the clearest evidence that the missing constraints, and not the tuning, separate the two implementations.

\subsection*{Truth-plant fidelity ladder and reproducibility boundary}

Every transfer campaign in Section~\ref{sec:results} is flown on the simulator of Table~\ref{tab:sim}, with every environment term switched on and vehicle parameters redrawn per seed, and against the error model of Table~\ref{tab:errors}. The configuration is recorded in the metadata of every output record, and all $5000$ records behind Section~\ref{sec:results} carry it, so no result in this paper mixes configurations. Truth fidelity is also distinct from the model inside the optimizer: changing the propagation does not turn a YA or element-space controller into nonlinear MPC. The full translational truth of the V2 rungs is
\begin{equation}
\dot{\mathbf v}=\mathbf a_{50c}+\mathbf a_{3B}+\mathbf a_{\rm SRP}+\mathbf a_D,
\end{equation}
where the complete normalized degree/order-50 Mars50c potential is evaluated in a rotating Mars-fixed frame and transformed back to inertial coordinates. Solar third-body gravity is differential, SRP includes cylindrical eclipse, and drag uses a co-rotating exponential atmosphere. The attitude rung adds $I\dot{\boldsymbol\omega}=\boldsymbol\tau-\boldsymbol\omega\times I\boldsymbol\omega$ and quaternion kinematics, with gravity-gradient, SRP and drag torques. Which of these numbers the reference publishes and which are ours is marked term by term in the last column of Table~\ref{tab:sim}, so the boundary between reproduction and assumption is visible without consulting the source.

Table~\ref{tab:errors} states the navigation model in the units it is drawn in rather than as a citation: the published $3\sigma$ bounds divided by three, split per component, and interpolated linearly in range between $100$~m and $5$~km. One entry needed interpretation, because the source labels every column a percentage of range and that is dimensionally impossible for velocity; beyond $5$~km the velocity figure is therefore read as the absolute value printed. The simulator sends every command through one physical actuator path that limits net support to $20$~N, allocates the assumed eight-thruster block, applies the 1~s pulse train, MIB, quantisation and on-time error, realizes the state impulse and charges the same individual on-times. This removes the V1 separation between an ideal state jump and a fuel-only derating.

\begin{table*}[tbp]
\centering
\caption{The truth simulator every campaign is flown on. The last column separates what Ref.~\cite{Hartley2012} publishes from what we assume.}
\label{tab:sim}
\small
\setlength{\tabcolsep}{4pt}
\begin{tabularx}{\textwidth}{@{}p{0.15\textwidth}p{0.24\textwidth}Xl@{}}
\toprule
Effect & Model & Parameters & Provenance \\
\midrule
Central gravity & point mass & $\mu_{\mathrm M}=4.283465\times10^{13}$~m$^3$s$^{-2}$ & Mars50c \\
Non-spherical gravity & normalised $50\times50$ field, evaluated Mars-fixed and rotated back & $R_{\mathrm M}=3396.19$~km, $J_2=1.960454\times10^{-3}$, $\omega_{\mathrm M}=7.0882\times10^{-5}$~rad\,s$^{-1}$ & Mars50c; prime-meridian epoch assumed $0^\circ$ \\
Third body & Sun, differential & $\mu_\odot=1.327124\times10^{20}$~m$^3$s$^{-2}$, $a_{\mathrm M}=1.5237$~AU, $e_{\mathrm M}=0.0934$ & standard \\
Solar radiation pressure & cannonball with cylindrical shadow, so eclipse entry and exit are modelled & $S_{1\mathrm{AU}}=1361$~W\,m$^{-2}$, $A=10$~m$^2$, $C_r=1.3$ & $A$, $C_r$ assumed \\
Atmospheric drag & co-rotating exponential atmosphere & $\rho_0=0.020$~kg\,m$^{-3}$ at the surface, $H=11.1$~km, $C_d=2.2$, $A=10$~m$^2$ & assumed \\
Attitude & rigid body, $I\dot{\boldsymbol\omega}=\boldsymbol\tau-\boldsymbol\omega\times I\boldsymbol\omega$, with quaternion kinematics & gravity-gradient, SRP and drag torques; centre-of-pressure offsets $\pm0.05$~m on body $x$ & inertia published; offsets assumed \\
Actuation & $8\times22$~N block, $20$~N guaranteed net translation, $1$~s pulse train & minimum impulse bit $0.068$~s, on-time quantisation $0.01$~s, on-time error $1\sigma=0.01$~s & published \\
Integration & RK4 & $10$~s maximum translational step, $2$~s attitude step & --- \\
\bottomrule
\end{tabularx}
\end{table*}

The reference scenario is the MSRE approach of Table~\ref{tab:phases}, driven by one configuration shared by the simulations, Monte Carlo runs, and figures. Table~\ref{tab:requirements} lists the orbit and vehicle parameters and Table~\ref{tab:errors} every error acting on them. Initial elements are uniform on their ranges, $\delta a\in[-50,50]$~km and $\delta i,\delta\Omega,\delta\varpi\in[-0.3,0.3]^\circ$ for the chaser, $\delta a\in[-10,10]$~km and $\delta\nu\in[0,320]^\circ$ for the target. The tube itself is designed against the per-step bounds $\bar{\mathbf w}^{\mathrm{tr}}=[5~\mathrm{m}^{\times3},5\times10^{-3}~\mathrm{m/s}^{\times3}]$ in the transfer phases and $\bar{\mathbf w}^{\mathrm{fin}}=[0.05,5\times10^{-4}]$ in the final one, rescaled with $T_s$ as $(T_s/T_s^{\mathrm{ref}})^2$ on position and $(T_s/T_s^{\mathrm{ref}})$ on velocity, since a bounded acceleration error moves velocity linearly in the step and position quadratically. Holding the width fixed while shortening $T_s$ empties the tightened input box and the vehicle free-drifts, so the rescaling is not optional. Draws are paired: for a given seed every controller sees the same dispersion, the same navigation stream and the same actuator execution error, so a comparison is a median of per-trial differences rather than a difference of medians between independent samples.

\begin{table}[tbp]
\centering
\caption{Scenario and vehicle parameters. The state box is inert as flown.}
\label{tab:requirements}
\begin{tabular}{@{}lc@{}}
\toprule
Parameter & Value \\
\midrule
Semi-major axis $a$ & 4\,643 km \\
Eccentricity $e$ & 0.204 \\
Orbit period & 2.67 h \\
Chaser mass & 1\,575 kg \\
Guaranteed net thrust $F_c$ & 20 N \\
Thruster block & $8\times22$ N, MIB 0.068 s \\
Approach & 300 km $\to$ contact \\
State box (as flown) & inert, $\pm10^{9}$ m \\
Capture requirement & $\sqrt{y^2+z^2}<0.20$ m \\
Blinding range $d_b$ & 3 m, then free drift \\
\bottomrule
\end{tabular}
\end{table}

\begin{table*}[tbp]
\centering
\caption{Every error the vehicle sees. Navigation and execution act at every guidance call; the mismatch scales are drawn once per seed.}
\label{tab:errors}
\small
\begin{tabular}{@{}llc@{}}
\toprule
Source & $1\sigma$, per axis & worst realised \\
\midrule
\multicolumn{3}{@{}l}{\emph{Navigation, zero mean, redrawn every guidance call}} \\
\quad range $<100$~m & $8.25\times10^{-3}$~m; $(3.00,2.33,2.33)\times10^{-3}$~m/s & --- \\
\quad $100$~m to $5$~km & linear in range to $0.577$~m; $1.92\times10^{-2}$~m/s & --- \\
\quad range $>5$~km & $1.92\times10^{-4}\rho$; $5.77\times10^{-4}$~m/s & --- \\
\midrule
\multicolumn{3}{@{}l}{\emph{Thruster execution, on every command}} \\
\quad minimum impulse bit & $0.068$~s $=9.50\times10^{-4}$~m/s, below which a command is dropped & --- \\
\quad on-time quantisation & $0.01$~s $=1.40\times10^{-4}$~m/s & --- \\
\quad on-time error & $0.01$~s $=1.40\times10^{-4}$~m/s & --- \\
\midrule
\multicolumn{3}{@{}l}{\emph{Parameter mismatch, multiplicative, one draw per seed}} \\
\quad Area & $0.10$ & $25.4\%$ \\
\quad Drag coefficient $C_d$ & $0.10$ & $24.5\%$ \\
\quad Reflectivity $C_r$ & $0.10$ & $21.6\%$ \\
\quad Mass & $0.05$ & $17.9\%$ \\
\quad Inertia & $0.05$ & $11.1\%$ \\
\quad Thrust & $0.02$ & $4.6\%$ \\
\quad Gravitational $\mu$ & $10^{-5}$ & --- \\
\quad Initial pointing & $2^\circ$ about the pointing solution & $5.3^\circ$ \\
\quad Initial body rate & $0.05^\circ$/s & --- \\
\bottomrule
\end{tabular}
\end{table*}

\subsection*{Sharing one actuator between translation and pointing}

Thrust acts through the centre of mass, so a burn produces no torque and the two channels do not compete for momentum. They compete for on-time. Charging attitude work through the same propellant budget at $\Delta v_{att}=c\|\boldsymbol\tau\|$, with $c=2t_{\mathrm{burn}}/(\ell m)$ set by the lever arm and the pulse length, makes that competition explicit and lets it be written as one constraint per step,
\begin{equation}\label{eq:couple}
|v_{k,i}|+c\|\boldsymbol\tau_k\|_1\le u_{\mathrm{tight},i,k},
\end{equation}
added to the final-phase program alongside pointing terms $\mathbf q_{v,k}^\top Q_q\mathbf q_{v,k}+\boldsymbol\omega_k^\top Q_\omega\boldsymbol\omega_k$ in the cost and $\|\boldsymbol\tau_k\|_\infty\le\tau_{\lim}/\sqrt3$. The $1$-norm in \eqref{eq:couple} upper-bounds the $2$-norm the charge is taken on, so the reservation never under-books. The attitude prediction that goes with it is linear: with $\boldsymbol\xi=[\mathbf q_{e,v};\boldsymbol\omega_e]$ the small-angle kinematics give a nilpotent $A_c$, for which forward Euler and the matrix exponential coincide, and the gyroscopic term $\boldsymbol\omega\times J\boldsymbol\omega$ vanishes to first order about $\boldsymbol\omega_e=\mathbf 0$. Only the predictor is linear. The plant integrates the full nonlinear rigid body with gravity-gradient and radiation-pressure torques, so the linearisation is a modelling error the loop absorbs rather than a change to the physics. The alternative form, labelled Proposed (uncoupled) below, runs the identical translation guidance with an uncoupled quaternion-feedback PD, $\boldsymbol\tau=-k_p\mathbf q_e-k_d\boldsymbol\omega_e$ saturated at $\tau_{\lim}$, executed after the translation program has already committed the interval so that neither sees the other.

\section{Results}\label{sec:results}

\subsection*{Two tests, and why they must be reported apart}

Every comparison in this section is a full transfer from $300$~km to capture on the simulator of Table~\ref{tab:sim}, $500$ dispersed draws per scenario per form, seeds paired across forms by construction. Control stops at $\rho=3$~m and the vehicle coasts, so the arrival is scored on a free-drift crossing of the target plane rather than on a commanded state, and that scoring involves two questions which earlier drafts of this paper welded into one phrase. The first is whether the coast crosses the plane at all. The second is whether the crossing falls inside the $0.20$~m requirement circle. A draw can pass the first and fail the second, and on the elliptical scenario up to $48$ draws in $500$ do exactly that, so Table~\ref{tab:headtohead} prints the two counts in separate columns and no sentence below merges them.

Requirement compliance is where the result is strongest and it is eccentricity-specific. On the circular scenario every form is at or within one draw of $500$ of $500$, and nothing separates them. On the elliptical scenario the proposed form is at $500$ of $500$ and the three replication variants are at $468$, $463$ and $452$, that is $93.6$, $92.6$ and $90.4\%$, with Wilson intervals of $[91.1,95.4]$, $[90.0,94.6]$ and $[87.5,92.7]\%$ that reach nowhere near the proposed form's $[99.2,100]$. Roughly one elliptical transfer in eleven arrives outside the requirement under the reimplemented benchmark, and none does under the tube. Three of the replication misses sit within a millimetre of the boundary and would flip under a looser reading of it, which changes no interval materially.

\subsection*{Propellant, paired seed by seed}

\begin{figure}[tbp]
\centering
\includegraphics[width=\columnwidth]{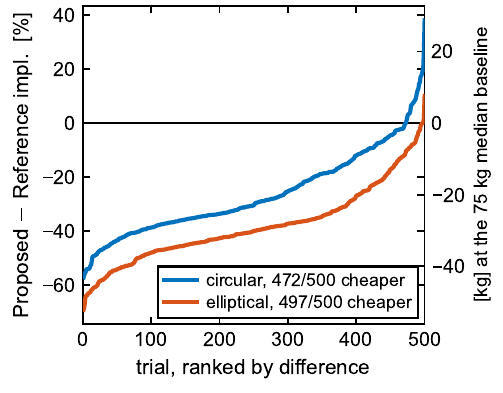}
\caption{Per-trial propellant difference against the best replication, sorted, as a percentage of that trial's own baseline. The right axis converts at one number and is a reading aid.}
\label{fig:paired}
\end{figure}

Propellant separates in both scenarios and by a margin the unpaired statistics would hide. The per-draw standard deviation is $17$ to $30$~kg against differences of $24$ to $33$~kg, so a difference of means is nearly uninformative while the paired difference is decisive. Against the best replication the proposed form is $24.05$~kg cheaper on the circular scenario, winning $472$ of $500$ matched seeds, and $31.39$~kg cheaper on the elliptical one, winning $497$ of $500$ (Table~\ref{tab:headtohead}, Fig.~\ref{fig:paired}). Expressed against the baseline those are savings of $29.2$ and $39.6\%$; the same numbers read the other way round, as the baseline spending more than the proposed form does, are $41.3$ and $65.4\%$, and the two differ only in the denominator. Mean totals are $58.18$ and $47.97$~kg against $82.23$ and $79.37$. The ranking is unchanged across all three replication variants, which span $81.85$ to $85.56$~kg circular and $79.37$ to $80.94$ elliptical, so the comparison does not rest on picking a convenient baseline.

\begin{figure*}[tbp]
\centering
\includegraphics[width=0.92\textwidth]{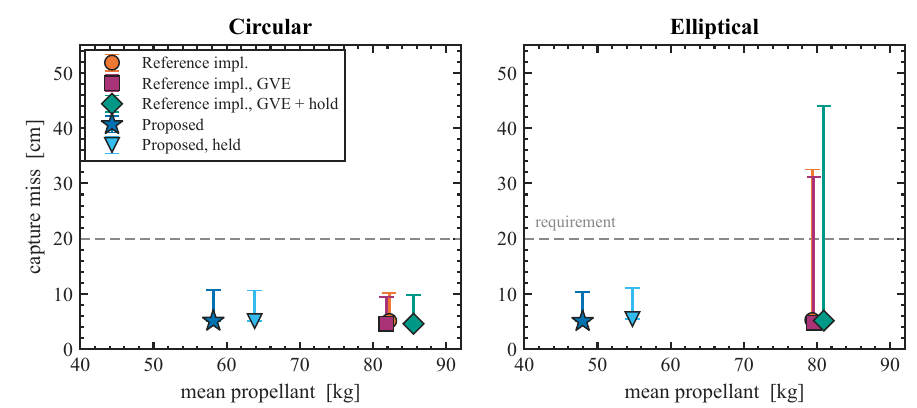}
\caption{The ten cells of Table~\ref{tab:headtohead}, propellant against capture accuracy. Marker at the median miss, whisker to the $95$th percentile. The reimplementation family collapses onto a single propellant cluster under dispersion on both scenarios, and its eccentric tails cross the requirement while neither proposed cell does.}
\label{fig:fuelmiss}
\end{figure*}

Nothing about this is free. The proposed form flies $15.41$ and $9.03$ orbits against $5.55$ and $4.47$, factors of $2.8$ and $2.0$, and a mission that cannot spend the extra time cannot spend the saving either. The replication front-loads its correction and is finished inside six orbits; the tube spreads the same correction over a longer coast and never reaches the replication's level.

\subsection*{The tail, and the one draw that runs the other way}

\begin{figure*}[tbp]
\centering
\begin{minipage}[t]{0.49\textwidth}\centering
\includegraphics[width=\linewidth]{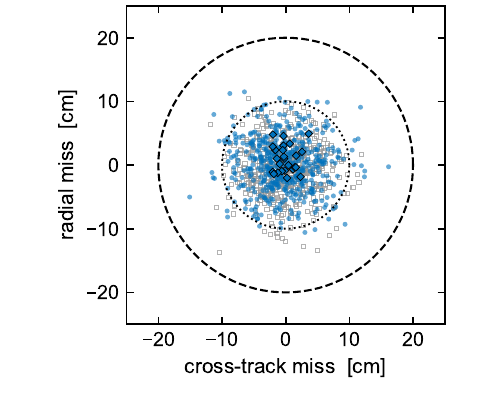}\\
{\small (a) circular}
\end{minipage}\hfill
\begin{minipage}[t]{0.49\textwidth}\centering
\includegraphics[width=\linewidth]{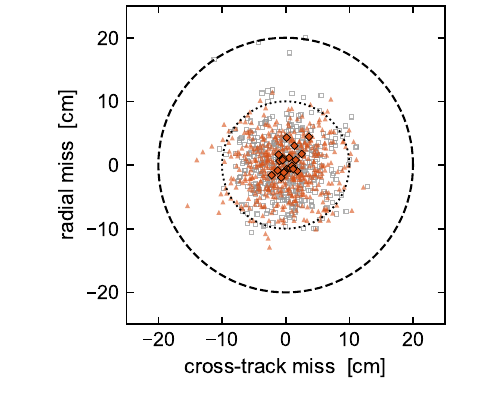}\\
{\small (b) elliptical}
\end{minipage}\par\vspace{3pt}
\includegraphics[width=0.94\textwidth]{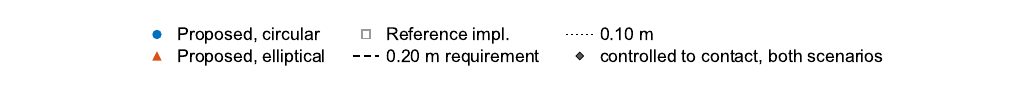}
\caption{Free-drift crossings of the target plane, $500$ draws per scenario, against the $0.20$~m requirement and a $0.10$~m inner ring. Both panels share the key beneath them.}
\label{fig:cap}
\end{figure*}

The tail splits by scenario in the same direction the requirement compliance does. On the elliptical scenario the proposed form holds a $95$th-percentile miss of $10.39$~cm against $32.55$ for the best replication and $43.97$ for the faithful reimplementation, factors of $3.1$ and $4.2$, and a worst case of $16.69$~cm against $170.01$ and $185.33$. On the circular scenario the ordering reverses and the margin is small, $10.73$~cm against $10.18$ and $9.82$, and the medians are within $0.2$~cm across every cell, so median accuracy discriminates nothing and is not claimed to. Figure~\ref{fig:cap} draws the two clouds against the requirement. A draw that misses widely leaves the frame rather than landing at its edge: one proposed draw does so on the circular scenario and none of the reference, and on the elliptical one $29$ reference draws do so and no proposed one. The diamonds are the same guidance flown to the interface under control instead of released at $3$~m, read as the lateral offset where control ends, because a vehicle still being controlled has no drift to score; over the $25$ and $20$ cases of the arrival bank the medians are $2.22$ against $3.36$~cm circular and $1.42$ against $5.77$~cm elliptical, so the accuracy floor is set by the uncontrolled last three metres and not by the guidance.

One circular draw reaches $179.17$~cm and is reported rather than trimmed. It is the only proposed-form arrival outside the requirement in $1000$ draws, and it does not originate in the final approach: its hand-over lateral rate is $13$ times the $99.8$th percentile of the same campaign, and across the $500$ circular draws that rate correlates with the eventual miss at $r=0.963$. Excluding it the circular worst case is $16.16$~cm. The draw arrives at the final approach already broken, which is a hand-over gate question rather than a guidance one, and the gate that would catch it is the in-track and orbit-match conjunction of Section~\ref{sec:phased} rather than anything inside \eqref{eq:qp}.

Isolating the attitude coupling makes the tail argument concrete, and it is a smaller experiment than the rest of this section. Proposed (uncoupled) is the identical guidance with the uncoupled attitude PD of Section~\ref{sec:setup}, run against Proposed on $25$ paired draws per scenario at the same fidelity, and the two must be read only against each other because $25$ draws will not support a comparison with the $500$-draw cells above. On those draws the coupling costs $0.02$~kg, which is nothing, and moves the $95$th-percentile miss from $10.87$ to $8.70$~cm circular and from $12.18$ to $7.94$~cm elliptical while leaving the median half a centimetre worse. Solving translation and pointing in one program buys the tail and nothing else, which is what \eqref{eq:couple} predicts, since attitude is one to three per cent of the phase charge and that fraction is the ceiling on what the coupling can be worth in propellant.

\begin{table*}[tbp]
\centering
\caption{Propellant separates in both scenarios; compliance and the tail only in the elliptical one. $500$ dispersed draws per cell on paired seeds. The last two columns are the paired comparison against the proposed form: a positive median difference means the baseline spends more on the median seed, and \emph{cheaper} counts the seeds on which it does.}
\label{tab:headtohead}
\setlength{\tabcolsep}{4.5pt}
\begin{tabular}{@{}ll cc ccc rrr rc@{}}
\toprule
Form & Scen. & Cross & Within & \multicolumn{3}{c}{miss [cm]} & \multicolumn{3}{c}{propellant} & \multicolumn{2}{c}{paired vs prop.} \\
\cmidrule(lr){5-7}\cmidrule(lr){8-10}\cmidrule(lr){11-12}
 & & /500 & /500 & med & p95 & worst & [kg] & vs ref. & vs prop. & med $\Delta$ & cheaper \\
\midrule
Ref. impl. (Kep.) & C & 500 & 500 & 5.14 & 10.18 & 17.25 & 82.23 & -- & $+41.3$ & $+25.20$ & 472/500 \\
Ref. impl. (GVE) & C & 500 & 500 & 4.67 & 9.47 & 14.45 & 81.85 & $-0.5$ & $+40.7$ & $+24.82$ & 471/500 \\
Ref. impl. (GVE, held) & C & 500 & 500 & 4.63 & 9.82 & 14.45 & 85.56 & $+4.0$ & $+47.0$ & $+28.87$ & 479/500 \\
\textbf{Proposed} & \textbf{C} & \textbf{500} & \textbf{499} & \textbf{5.15} & \textbf{10.73} & \textbf{179.17} & \textbf{58.18} & $-29.2$ & -- & -- & -- \\
Proposed (held) & C & 500 & 500 & 5.11 & 10.63 & 16.01 & 63.84 & $-22.4$ & $+9.7$ & $+5.46$ & 395/500 \\
\midrule
Ref. impl. (Kep.) & E & 499 & 468 & 5.31 & 32.55 & 170.01 & 79.37 & -- & $+65.4$ & $+33.62$ & 497/500 \\
Ref. impl. (GVE) & E & 500 & 463 & 4.82 & 31.17 & 185.33 & 79.58 & $+0.3$ & $+65.9$ & $+34.19$ & 495/500 \\
Ref. impl. (GVE, held) & E & 500 & 452 & 5.19 & 43.97 & 185.33 & 80.94 & $+2.0$ & $+68.7$ & $+34.93$ & 499/500 \\
\textbf{Proposed} & \textbf{E} & \textbf{500} & \textbf{500} & \textbf{5.10} & \textbf{10.39} & \textbf{16.69} & \textbf{47.97} & $-39.6$ & -- & -- & -- \\
Proposed (held) & E & 500 & 499 & 5.48 & 11.13 & 36.73 & 54.78 & $-31.0$ & $+14.2$ & $+5.28$ & 405/500 \\
\bottomrule
\end{tabular}

\end{table*}

\begin{figure*}[tbp]
\centering
\includegraphics[width=0.94\textwidth]{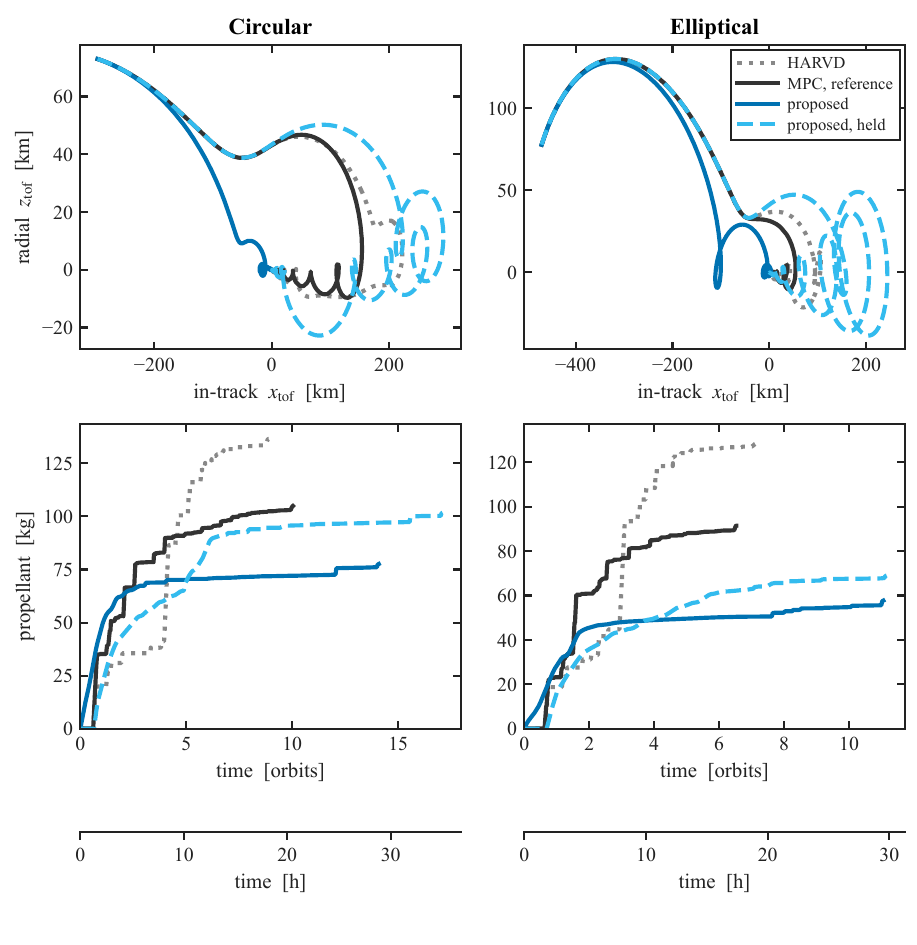}
\caption{The Fig.~$12$ conditions of Ref.~\cite{Hartley2012}, both scenarios, undispersed. Track above, charged propellant below, on the layout the reference uses. The held variant releases at the reference's own first-burn epoch and recovers the published geometry the released law does not, crossing to positive in-track and looping back instead of approaching monotonically; that detour is what its extra propellant buys, $101.8$ against $77.6$~kg circular and $69.0$ against $57.7$~kg elliptical.}
\label{fig:nominal}
\end{figure*}

\begin{figure*}[tbp]
\centering
\begin{minipage}[t]{0.49\textwidth}\centering
\includegraphics[width=\linewidth]{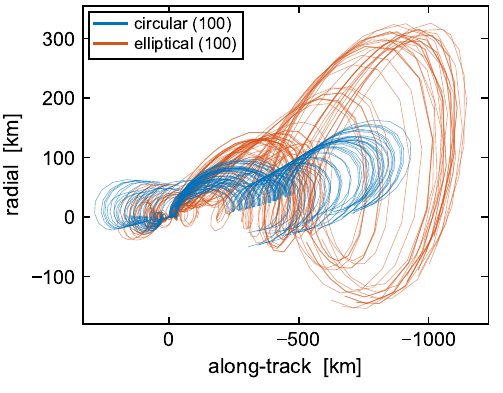}\\
{\small (a) trajectories}
\end{minipage}\hfill
\begin{minipage}[t]{0.49\textwidth}\centering
\includegraphics[width=\linewidth]{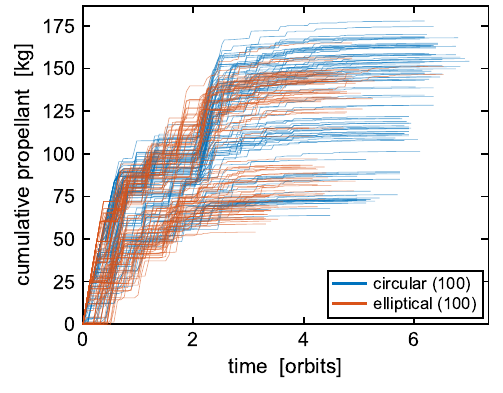}\\
{\small (b) propellant}
\end{minipage}
\caption{The initial-condition ensemble of Ref.~\cite{Hartley2012}, $224$ cases per scenario, of which $100$ are drawn at an even stride. (b) is against orbits, not seconds.}
\label{fig:ensemble}
\end{figure*}

\subsection*{How much of the margin is the head start}

\begin{figure}[tbp]
\centering
\includegraphics[width=\columnwidth]{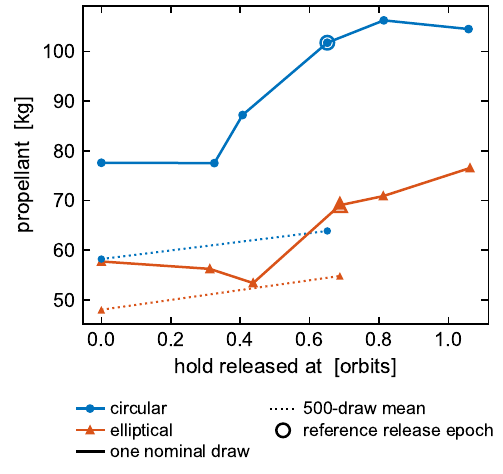}
\caption{Propellant against the epoch at which the hold releases. The abscissa is the released epoch, not the commanded one.}
\label{fig:wait}
\end{figure}

The faithful reimplementation reaches the published value partly because it holds its homing command before the first burn, so a reader is entitled to ask how much of the proposed form's margin is the controller and how much is firing earlier. Holding the proposed form to the same epoch answers it. The hold is commanded as a duration and released at the first guidance call at or after it, so the epoch flown is not the epoch asked for: against the measured reference release of $0.570$ orbits circular and $0.687$ elliptical, the homing command grid of one call per $600$~s releases at $0.651$ and $0.687$, the circular case landing $14.2\%$ late and the elliptical one happening to fall on its commanded value. What follows is therefore a slightly longer hold than intended on the circular scenario, and since a longer hold costs more, the cost quoted below is an upper bound on the cost of the epoch that was meant. The answer then depends on how many draws the question is asked over. On one nominal case the hold costs $31\%$ circular and $20\%$ elliptical, which would leave almost nothing of the circular margin. Over $500$ dispersed draws it costs $9.7$ and $14.2\%$, and the held form still spends $63.84$ and $54.78$~kg against $81.85$ to $85.56$ and $79.37$ to $80.94$ for the replications. The reversal is the point rather than an inconvenience: a single-seed experiment put the head start at a third of the effect and the dispersed one puts it at a tenth, and the two disagree by more than either is worth. Figure~\ref{fig:wait} draws both on one axes. The hold also removes the circular outlier, taking the worst case from $179.17$ to $16.01$~cm, at the price of a worse elliptical worst case, $36.73$ against $16.69$.

\subsection*{Where the propellant goes, and against what floor}

\begin{figure}[tbp]
\centering
\includegraphics[width=\columnwidth]{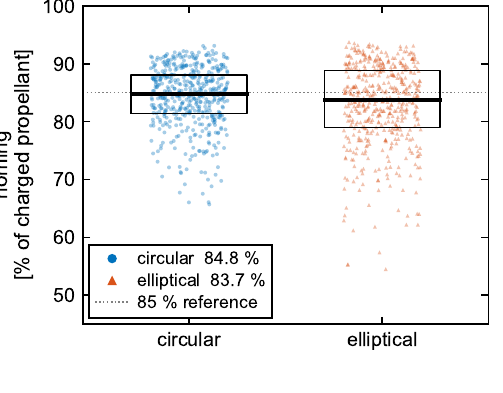}
\caption{Homing share of charged propellant, per trial, $500$ draws per scenario.}
\label{fig:phaseprop}
\end{figure}

The phase split decides how much any of this can be worth, and it is lopsided. Over the $500$ captured draws of the proposed form, homing takes $50.04$~kg of a charged $58.46$ on the circular scenario and $40.93$ of $48.16$ on the elliptical one, the closing $6.70$ and $5.38$~kg, and the final approach $1.72$ and $1.85$. In percentages that is $86/11/3$ and $85/11/4$, and Fig.~\ref{fig:phaseprop} shows the split holds trial by trial rather than only on the mean, with a median homing share of $85$ and $84\%$ and an interquartile range of four to ten points. Two consequences follow directly. Any claim that a final-approach retuning delivers a large propellant saving is bounded above by about $2$~kg whatever the tuning does, and the $15.8$~m/s the horizon change of Section~\ref{sec:horizon} recovered can only have come from homing, which is the phase whose horizon is being chosen.

An absolute yardstick keeps the comparison honest in the other direction. The idealised two-impulse transfer for the circular scenario, a Hohmann-class correction of the $50$~km semi-major-axis error plus the $0.3^\circ$ plane change, costs $38.9$~m/s, of which the plane change alone is $17.4$. The proposed form charges a mean $81.4$~m/s over the $500$ draws, a factor of $2.1$ above that floor, and the replication $116.1$~m/s, a factor of $3.0$. Neither is close to optimal and neither should be, since the floor assumes perfect knowledge, instantaneous impulses and no acceptance window while both controllers fly a dispersed vehicle through navigation error to a $20$~cm interface. Quoting it establishes that a third saved is a third off a number that still has room in it, not a claim of optimality.

\subsection*{Computational cost}

The enumeration in \eqref{eq:jn} is what the horizon rule costs, and Table~\ref{tab:compute} prices it on the stress grid where the two caps can be compared cell by cell. Doubling the cap from $25$ to $50$ candidates multiplies the wall clock per cell by $2.53$ and buys ten cells that were previously unflyable together with $15.8$~m/s on the eleven that were. That cost is paid in a phase sampled at $600$~s, so it is not a real-time constraint; the final approach, which is sampled at $3$~s, solves a fixed $15$-step horizon on six states with no enumeration at all. OSQP median solve times for the tightened program run a few milliseconds and the tightened bounds add about a millisecond to the median, with a long right tail from cold starts.

\begin{table}[tbp]
\centering
\caption{What the horizon enumeration costs, on the $21$-cell stress grid.}
\label{tab:compute}
\setlength{\tabcolsep}{4pt}
\resizebox{\columnwidth}{!}{
\begin{tabular}{@{}lcc@{}}
\toprule
Quantity & $n_{\max}=25$ & $n_{\max}=50$ \\
\midrule
Stress cells flown & 11/21 & 21/21 \\
Median $\Delta v$, cells both flew [m/s] & 141.7 & 122.1 \\
Median wall clock per cell [s] & 168 & 423 \\
Wall clock, paired ratio & 1.00 & 2.53 \\
\bottomrule
\end{tabular}
}
\end{table}

\section{Discussion}\label{sec:disc}

The divergence at $e^\star$ is kinematic. Kepler's second law fixes the periapsis-to-apoapsis rate ratio at $((1+e)/(1-e))^2$; uniform-period sampling gives each step the same wall-clock duration, but a periapsis step covers a much larger true-anomaly arc. The YA matrices inherit this asymmetry, and as $e$ grows the worst-case step dominates $\bar A$. The same asymmetry that makes the YA model necessary, namely the failure of CW on eccentric orbits, is what eventually bounds the tube built on it.

Operationally the threshold partitions the design space, but only once the sampling period is fixed with it. At $T_s=200$~s the constant-gain LTV tube is admissible up to $e\approx0.65$, which covers planetary capture, Moon-class and Earth LEO operations and leaves Mars a $3.2\times$ margin; at the $600$~s homing period the same design certifies only to $0.362$, and at $900$~s to $0.249$, so an orbit at $e=0.5$ passes at the fast rate and fails at the slow one. GTO at $e=0.73$ is outside the envelope at every period tested here and demands a review of the disturbance budget before flight. Eccentricities approaching $0.7$ are the natural domain of a richer ancillary formulation, such as a periodic gain $K(\nu)$ following the orbit or variable-rate sampling that shortens $T_s$ at periapsis, and the certificate identifies exactly where that extra complexity begins to earn its cost. The underlying question, whether a fixed gain survives control authority that varies periodically along the orbit, recurs outside rendezvous: magnetic detumbling faces the same structure, with the achievable torque direction set by the local field rather than by true anomaly \cite{IskenderDetumbling2026}.

Two limitations bound how far any of this generalises. The threshold is sufficient and not tight, since it assumes a constant ancillary $K$, a uniform $T_s$ and an additive box disturbance bound, and richer formulations such as a $\nu$-scheduled gain or an ellipsoidal tube should extend the envelope by a body-dependent constant without changing the structural form of the recursion. The propellant result, meanwhile, is bought with time of flight at a factor of two to three, so it belongs to a mission that can spend orbits rather than to one on a fixed arrival window. The corridor diagnostic that would show the tightening removing constraint violations, the cross-body envelope and the geostationary co-location study are all carried in the companion journal treatment \cite{IskenderAISR2026}, where the scenarios that exercise them can be given the space to be described properly.

\section{Conclusion}\label{sec:concl}

A box-tightened tube is sound only while the predictor it is built on errs by less than the disturbance bound it was certified against, and \eqref{eq:valid} makes that condition a number that depends on the sampling period as much as on the orbit: $26.4$~km of separation at $300$~s, $13.2$~km at $600$~s. The bound earns its place by predicting rather than describing. It says that the homing phase, which opens $23$ times outside its own limit and overruns $\bar w$ by factors of $497$ and $852$, must be re-posed, and that the closing phase, which sits a factor of $2.6$ inside it, cannot benefit from the same treatment. Both were then measured: element space restored the certificate to $15$ to $28\%$ margin, while a cylindrical frame $128$ times more accurate over the closing phase changed propellant by $0.09\%$.

Tightening also makes the prediction horizon part of what is certified, and that coupling is where the second result comes from. The variable horizon itself is the benchmark's, adopted here with its own weight and its own selection functional; what changes is that the terminal tolerance available at horizon $N$ is eroded by a tightening that grows with $N$, so reach and margin are traded inside one enumeration. Raising the search cap from $25$ to $50$ steps took a $21$-cell stress grid from $11$ flown approaches to $21$, cost $2.5$ times the wall clock, and was $15.8$~m/s cheaper on the eleven cells both caps flew. The failure it removed was invisible in a twenty-draw campaign and total in the twenty-fifth, and its boundary is a joint condition on semi-major-axis error and target true anomaly, so no single-variable sweep would have located it. The cap is still binding at $50$: over $71\,398$ guidance calls the selected horizon sits within one step of it $22\%$ of the time.

Over $500$ dispersed transfers per scenario from $300$~km to capture, at $50\times50$ Mars gravity with third-body, radiation-pressure, drag and attitude terms and per-seed parameter mismatch, the attitude-coupled form arrives inside the $0.20$~m requirement on $499$ and $500$ of $500$ draws, against $452$ to $468$ of $500$ for the three replication variants on the elliptical scenario, and spends $24.1$ and $31.4$~kg less on $472$ and $497$ of $500$ matched seeds. Requirement compliance, the tail and the propellant margin all split by eccentricity rather than uniformly: the $95$th-percentile miss is $10.39$~cm elliptical against $32.55$, but $10.73$ against $10.18$ circular, so what is claimed is propellant in both scenarios and compliance and tail in the elliptical one. Solving translation and pointing in one program is what buys the tail, since on a $25$-draw paired ablation the uncoupled pointing loop costs $0.02$~kg and opens the same percentile from $8.70$ and $7.94$~cm to $10.87$ and $12.18$. The saving costs $2.8$ and $2.0$ times the time of flight, and set against an idealised two-impulse floor of $38.9$~m/s for the circular case the proposed form charges $81.4$~m/s, $2.1$ times the floor, against the replication's $116.1$.

Two results run the other way and one experiment reverses itself under dispersion. The eccentricity certificate is not the single scalar earlier work took it for: it is insensitive to where the ancillary gain is designed and to the design weights, but it falls by a factor of $3.4$ across the sampling periods a phased approach schedules, so quoting it without a period would certify an $e=0.5$ orbit that loses the certificate above $368$~s. The tightening is also not where the propellant is, since $85$ to $86\%$ of the budget is spent in homing and under $4\%$ in the final approach, so what the tube buys on this scenario is the horizon rule of Section~\ref{sec:horizon} rather than a corridor margin the inert state box never asks for. And the head-start ablation measured at one seed put the initial hold at a third of the propellant difference where five hundred dispersed draws put it at a tenth, which is a reminder that a nominal case can rank two designs the wrong way round. Robust constraint satisfaction is the only guarantee proved here. Recursive feasibility is not, because the terminal ellipsoid that would deliver it makes the program a QCQP the solver cannot accept and the campaigns run without any terminal constraint at all, and asymptotic stability is not, because with no terminal set the DARE cost is not a Lyapunov function. What carried the result was the reformulation the validity bound pointed to, the horizon it is chosen against, and the cost form.

\section*{Acknowledgements}
The authors thank the OSQP developers \cite{Stellato2020}, whose solver underpins every result reported here.

\bibliography{A3_EllipticalTubeMPC}

\end{document}